\RequirePackage{ifpdf}

\documentclass[twocolumn,preprintnumbers,amsmath,amssymb,secnumarabic]{revtex4}
\usepackage{dcolumn}
\usepackage{bm}

 \ifpdf
  \usepackage[%
    pdftex,%
    colorlinks,%
    hyperindex,%
    plainpages=false,%
    bookmarksopen,%
    bookmarksnumbered%
  ]{hyperref}
  \usepackage{thumbpdf}
\else
  \usepackage{hyperref}
\fi

\usepackage{bm}%
\expandafter\ifx\csname package@font\endcsname\relax\else
 \expandafter\expandafter
 \expandafter\usepackage
 \expandafter\expandafter
 \expandafter{\csname package@font\endcsname}%
\fi \numberwithin{equation}{section}

\begin{document}

\title{Comments on quant-ph:0609176}%

\author{XiaoDi Wu}%
\email[Email me at ]{1suncat1@163.com}
 \affiliation{Department of Physics , Tsinghua
University, Beijing, China, 100084}

\begin{abstract}
In this note, we show the mistake which has been made in quant-ph
0609176. Further more, we provide a sketch of proof to show the
impossibility of the effort of such kind toward improving the
efficiency of Grover's Algorithm.
\end{abstract}

\maketitle

\section{The mistakes}
In quant-ph 0609176~\cite{qph01}, the author provides a kind of
quantum circuit using Toffoli Gate. The properties regarding such
circuit are provided and correctly analyzed in the original letter.
However, the authors fail to see some basic rules of quantum circuit
when they try to analysis their proposed algorithm for unsorted
database search problem.

For any circuit, the state is the tensor product of every qubit in
the circuit. Thus the analysis of quantum circuit can not be limited
within subsystem. If in this way, it is easy to make mistakes about
the superposition and entanglement properties of the whole process.
The following derivation will use the denotation in ~\cite{qph01}.

More precisely, in~\cite{qph01}, the proposed algorithm for
unsorted database search algorithm starts with the state\\
\begin{equation*}
    |\varphi\rangle=(\frac{1}{\sqrt{N}}\sum_{i=0,i\neq
    d}|x_i,0\rangle+ \frac{1}{\sqrt{N}}|x_d,1\rangle) \otimes |w_0\rangle^{\otimes
    M}
\end{equation*}

Only in the first round, there are operations on the first $n$ qubit
of the circuit. The first part of the state $|\varphi\rangle$ will
remain unchanged, namely $\frac{1}{\sqrt{N}}\sum_{i=0,i\neq
d}|x_i,0\rangle\otimes|w_0\rangle^{\otimes M}$. At the same time the
second part will become
$\frac{1}{\sqrt{N}}|x_d,1\rangle\otimes|w_d\rangle^{\otimes M}$. It
can be seen that in further rounds, the component of the state which
has $|w_d\rangle$ will stay in the second part and the first part
will always remain the same. Finally, the probability to get any
knowledge of $|w_d\rangle$ when we measure won't exceed
$\frac{1}{N}$.

The mistake made by the authors is that they ignore all the
$q_i^{(j)}$ are in tensor product with others and therefore are
correlated. This property makes the effect of superpositions is not
the way the authors thought in their letter.

\section{Sketch of Proof of the impossibility}
Further more, we can show some heuristic ideas of the question
whether extra qubits will be helpful to improve the efficiency for
algorithm of unsorted database problem.  Note that the method
in~\cite{qph01} is belonging to the extra qubits using type.

To show our idea, we need to see how the original Grover's Algorithm
works. The Grover's Algorithm starts with state $\varphi$ and
unitary operation $U$ and oracle relative operation $Q^T$ are
alternatively used. Therefore, the evolution of the algorithm can be
expressed in the following way.
\begin{equation} \label{equ:evolution}
    |\varphi\rangle=U_tQ^T_tU_{t-1}Q^T_{t-1}\dots
    U_1Q^T_1U_0|\psi\rangle
\end{equation}
Here $t$ refers to the step of time complexity of the algorithm.
Finally we will measure the state $|\varphi\rangle$. Assuming
$|\tau\rangle$ is the state we want, the probability we will get the
answer is $|\langle\tau|\varphi\rangle|^2$. Grover's algorithm shows
that in order to detect the answer with a constant probability, we
need to query the oracle $O(\sqrt{N})$ times. Because of the result
in~\cite{BBBV97} that for a Quantum Turing Machine the complexity
for unsorted database search is $\Omega(\sqrt{N})$, the Grover's
algorithm has reached the lower bound and is optimal in some sense.

Now, we consider the extra using of qubits. We denote the state now
as $|\varphi\rangle \otimes |w\rangle$ where $|\varphi\rangle$ is
the binary representation of the elements and $|w\rangle$ represents
for the auxiliary qubits(assuming length-m). The final state is in
the same form $|\psi\rangle=|\varphi\rangle \otimes |w\rangle$ where
$|\varphi\rangle$ and $|w\rangle$ share the same meaning with above.
It is easy to see we can always meet such requirement. If some
algorithm ends with another form of state, we can use a unitary
operation to transform it to the form we need.

Assume the goal state is $|\tau\rangle$,and the auxiliary qubits
which are valid to provide answer are in the set $\Omega=\{w_i\}$,
where $i\leq 2^m$. The probability we will get the answer
$|\tau\rangle$ is \\
\begin{equation}\label{equ:probability}
    Pr_{success}=\sum_{w_i \in \Omega}|\langle\psi|(|\tau\rangle\otimes|w_i\rangle)|^2
\end{equation}
If $|\langle\psi|(|\tau\rangle\otimes|w_i\rangle)|^2$ are almost the
same for each $i$, the formula above can be written in following
form\\
\begin{equation}\label{equ:pro_1}
    Pr_{success}=|\Omega||\langle\psi|(|\tau\rangle\otimes|w_i\rangle)|^2\leq
    2^m|\langle\psi|(|\tau\rangle\otimes|w_i\rangle)|^2
\end{equation}

As we can see in~\ref{equ:pro_1}, if we can use all possible result
of the auxiliary qubits, the equality holds. In this situation, the
condition $|\langle\psi|(|\tau\rangle\otimes|w_i\rangle)|^2$ is
$O(2^{-m})$ suffices to guarantee the constant probability of
$Pr_{success}$. If in that situation, in order to achieve the
successful probability $2^{-m}$ of finding the answer in $2^{n+m}$
database we need to use $O(\arcsin{(\sqrt{2^{-m}})}\sqrt{2^{n+m}})$.
For large $m$, this result is near $O(2^{\frac{n}{2}})=O(\sqrt{N})$.
Namely, no improvement of the original algorithm. Note that here we
do not propose a detail or feasible algorithm, while we just propose
a necessary condition for this kind of algorithm.

Finally, we will provide reasonable argument to show that the
situation above is almost the case. First, if the probability is
condensed to certain kinds of auxiliary qubits, namely $|\Omega|$ is
much smaller than $2^m$, to achieve the constant probability of the
result, it is required that
$|\langle\psi|(|\tau\rangle\otimes|w_i\rangle)|^2=2^{-p}$ is much
larger than $2^{-m}$ for certain $w_i$. Because $|\tau\rangle$ is
uniformly distributed, we need to run
$O(\arcsin{(\sqrt{2^{-p}})}\sqrt{2^{n+m}})=O(2^{\frac{n+m-p}{2}})$
times which is not efficient. In this case, we can use Hadamard
transform on the auxiliary qubits to average the probability and
finally the case will be changed into the situation in our
derivation.


\end{document}